\documentclass[12pt]{article}
\usepackage{epsfig}
\input epsf

\def\gappeq{\mathrel{\rlap {\raise.5ex\hbox{$>$}}
{\lower.5ex\hbox{$\sim$}}}}
\def\lappeq{\mathrel{\rlap{\raise.5ex\hbox{$<$}}
{\lower.5ex\hbox{$\sim$}}}}

\newcommand{\be}{\begin{equation}}
\newcommand{\ee}{\end{equation}}
\newcommand{\ba}{\begin{eqnarray}}
\newcommand{\ea}{\end{eqnarray}}
\newcommand{\bi}{\begin{itemize}}
\newcommand{\ei}{\end{itemize}}

\def\lsi{\raise0.3ex
\hbox{$<$\kern-0.75em\raise-1.1ex\hbox{$\sim$}}}
\def\gsi{\raise0.3ex
\hbox{$>$\kern-0.75em\raise-1.1ex\hbox{$\sim$}}}

\begin{document}
\topmargin -1.0cm
\oddsidemargin -0.8cm
\evensidemargin -0.8cm
\pagestyle{empty}
\begin{flushright}
UNIL-IPT-01-2\\
hep-th/0102161\\
February 2001
\end{flushright}
\vspace*{5mm}

\begin{center}
{\Large\bf Extra dimensions as an alternative to Higgs mechanism?}

\vspace{1.0cm}

{\large 
Mikhail Shaposhnikov$^{a,}$\footnote{E-mail:
Mikhail.Shaposhnikov@ipt.unil.ch}
and Peter Tinyakov$^{a,b,}$\footnote{E-mail: Peter.Tinyakov@cern.ch}}\\
\vspace{.6cm}
{\it $^a$Institute of Theoretical Physics, University of Lausanne,\\
CH-1015 Lausanne, Switzerland\\
$^b$Institute for Nuclear Research, 117312 Moscow, Russia}\\

\vspace{.4cm}
\end{center}

\vspace{1cm}
\begin{abstract}
We show that a pure gauge theory in higher dimensions may lead to an
effective lower-dimensional theory with massive vector field, broken
gauge symmetry and no fundamental Higgs boson. The mechanism we
propose employs the localization of a vector field on a
lower-dimensional defect. No non-zero expectation values of the
vector field components along extra dimensions are required. New
possibilities for the solution to the gauge hierarchy problem are
discussed.
\end{abstract}

\vfill

\eject
\pagestyle{empty}
\setcounter{page}{1}
\setcounter{footnote}{0}
\pagestyle{plain}

It is usually believed that fundamental or composite scalar fields are
necessary for construction of self-consistent theory of interacting
massive vector bosons. Indeed, the only known renormalizable theories
of massive vector bosons in 4d are gauge theories with spontaneous
symmetry breaking by the Higgs mechanism, while all theories that have
good high energy behaviour of tree amplitudes can be shown
\cite{Cornwall} to be equivalent to the spontaneously broken gauge
theories with scalar fields.

In this note we show that a higher-dimensional pure gauge theory may
lead to an effective low-energy theory of {\em massive} vector bosons
in lower dimensions, thus breaking the gauge symmetry without
fundamental Higgs boson. Our mechanism makes use of the localization
of a vector field on a lower-dimensional defect (for localization of
massless vector fields on a brane see \cite{Dvali:1997xe} -
\cite{Dimopoulos:2000ej}). It does not involve scalar fields and does
not assume non-zero expectation values of the vector-field components
along extra dimensions. It is also quite different from the Hosotani
mechanism \cite{Hosotani} related to a non-trivial topology of the
compact extra dimensions.

In the usual Kaluza-Klein (KK) compactification, the low-energy sector
consists of the fields which are constant along the compact dimensions
and correspond to the lowest KK mode. Each higher-dimensional vector
field produces a massless vector and massless scalars (the vector
components in the direction of the compact dimensions) in the
low-dimensional effective theory. Higher KK modes give massive
fields. The gauge symmetry of the higher-dimensional theory with
respect to gauge transformations which do not depend on the compact
dimensions translates into the gauge symmetry of the effective
theory. The massive KK fields are invariant under this symmetry.

The mechanism we propose leads to a theory which is similar in many
respects. The crucial difference is that the analog of the massless KK
sector disappears from the theory, while massive sectors are organized
in such a way that one is much lighter than the others.  As a result,
the effective low-energy theory contains only a massive vector field
and has no gauge symmetry. 

Consider for definiteness the following gauge-invariant Lagrangian in
5d (generalizations to higher dimensions are obvious): \be S = -
\frac{1}{4} \int d^4x dz \Delta(z)F_{AB}F^{AB}~, \label{action} \ee
where $F_{AB}$ is the ordinary field strength and $\Delta(z)>0$ is
some weight function depending, in general, on the fifth coordinate
$z$. The origin of the weight function is not essential in what
follows. For instance, it may arise in a natural way from warped
compactifications on 3-branes \cite{RS2,seif,rusu}.  For simplicity,
we assume that $\Delta(z)$ is an even function of $z$. In the case
$\Delta(z) =1$ we have an ordinary 5d action. We are going to show
that, depending on the behaviour of $\Delta(z)$ at infinity, the 
low-energy effective theory may describe massive 4-dimensional
vector bosons without any scalar fields.

Let us consider first an Abelian gauge field. One can expand the field
$A_B(x_\nu,z)$ in a Fourier-type series with respect to the fifth
coordinate,
\be
A_B(x_\nu,z) = \sum_n A_B^n (x_\nu)\psi_n(z). 
\ee
From equations of motion for the gauge field it follows that the
functions $\psi_n(z)$ obey the equation
\be
-\frac{1}{\Delta(z)}\frac{\partial}{\partial z}\left(\Delta(z)
\frac{\partial}{\partial z}\psi_n(z)\right) = m_n^2\psi_n(z)
\label{main}
\ee
and the following orthogonality and completeness conditions:
\be
\int dz \Delta(z)\psi_n(z)\psi_m(z) = \delta_{mn},~~\sum_n\psi_n(z)
\psi_n(z')=\frac{1}{\Delta(z)}\delta(z-z').
\label{oreq}
\ee
From a 4-dimensional point of view, $A_{\mu}^n (x_\nu)$ are vector
fields, while $A_z^n (x_\nu)$ are scalars (here and below the Greek
indices run from 0 to 3 and correspond to 4-dimensional space-time).

The eigenvalues $m^2_n$ determine masses of the fields. As follows
from eq.(\ref{main}) upon multiplication by $\Delta(z)\psi_n(z)$ and
integration over $z$, they are all non-negative.  In the gauge $A_z=0$
the Lagrangian reads
\be
L = \sum_n\left(-\frac{1}{4} F^n_{\mu\nu}F_n^{\mu\nu} + \frac{1}{2}
m_n^2 (A^n_\mu)^2 \right)~,
\label{Leff}
\ee
where 
\be
F^n_{\mu\nu}=\partial_\mu A^n_\nu-\partial_\nu A^n_\mu~.
\label{}
\ee
Note that at $m^2_0\neq 0$, this Lagrangian describes a collection of
massive vector fields, while the scalar components have disappeared
from the spectrum. The counting of number of degrees of freedom works
in a trivial way: massless vector field in 5d has the same number of
physical degrees of freedom as the massive vector field in 4d.

If the fields corresponding to $n>0$ are much heavier, $m^2_n \gg
m^2_0$, one may expect that they decouple and eq.(\ref{Leff})
truncated at $n=0$ is the correct effective Lagrangian at low
energies. Thus, starting from the gauge-invariant 5-dimensional theory
(\ref{action}) we arrive at the 4-dimensional effective theory
(\ref{Leff}) containing a massive vector field and no scalars.

Let us now show that the desired situation --- non-zero $m^2_0$
separated by a gap from the rest of the eigenvalues --- occurs for a
wide class of weight functions $\Delta(z)$. The first thing to note is
that eq.(\ref{main}) always has a solution $\psi(z) = {\rm const}$ at
zero $m^2$. If the integral $\int dz \Delta(z)$ is convergent, the low
energy effective theory contains massless 4-dimensional vector field
(in other words, the gauge field is localized in 4 dimensions).  This
is analogous to the gravitational localization in the case of a local
string-like defect in 6d \cite{gs,baj,oda,Dubovsky} or a domain wall
in 5d with a dilaton field \cite{Kehagias}. Weight functions of that
type could be used to describe unbroken gauge symmetry on the brane.
Depending on the specific form of $\Delta(z)$, the theory may contain
a continuum of states living in the bulk without a mass gap, as in
the cases of refs.\cite{oda,Dubovsky,Kehagias}. For other choices of
the weight function the spectrum of states may develop a mass gap or
be discrete. For example, for $\Delta(z)=\exp(-M|z|)$ with $M > 0$ the
zero mode is separated from continuum by a mass gap $M$, whereas for
$\Delta(z)=\exp(-M^2 z^2)$ the spectrum of states is discrete with
$m_n^2 \sim M^2 n$. 

If the integral $\int dz \Delta(z)$ diverges, the solution $\psi(z) =
{\rm const}$ is non-normalizable. The linearly independent solution to
eq.(\ref{main}) with $m^2=0$ has the explicit form $\psi(z)= \int^z dx
\Delta^{-1}(x)$. Neither the second solution, nor any linear
combination with the first one is normalizable. Thus, in this case
zero mode does not exist and $m_0^2$ is positive. This is the
situation we are interested in.

If the weight function $\Delta(z)$ decreases at small $z$, reaches a
minimum and then grows fast enough at large $z$ (see Fig.~1a), the
lowest eigenvalue $m^2_0$ is generically small. In order to see this
we first rewrite eq.(\ref{main}) in the Schr\"odinger form by changing
the variables according to $\chi=\Delta^{1/2}\psi$,
\be
\left(-\frac{d^2}{dz^2} + V(z)\right)\chi_n(z) = m_n^2\chi_n(z)
\label{sch}
\ee
with the potential
\be
V(z)= W^2 - W',~~~ W= -{\Delta'\over 2\Delta}~.
\label{pot}
\ee 
For the weight function of Fig.~1a the potential is shown in
Fig.~1b; it has a well at small $z$ which gives rise to the lowest
level with almost zero eigenvalue. If $\Delta(z)$ grows exponentially
or faster at infinity, $V(z)$ goes to a constant or grows as well, and 
the spectrum has a gap or even is discrete. 
\begin{figure}[ht]
\begin{picture}(500,160)(0,0)
\put(10,10){
\epsfig{file=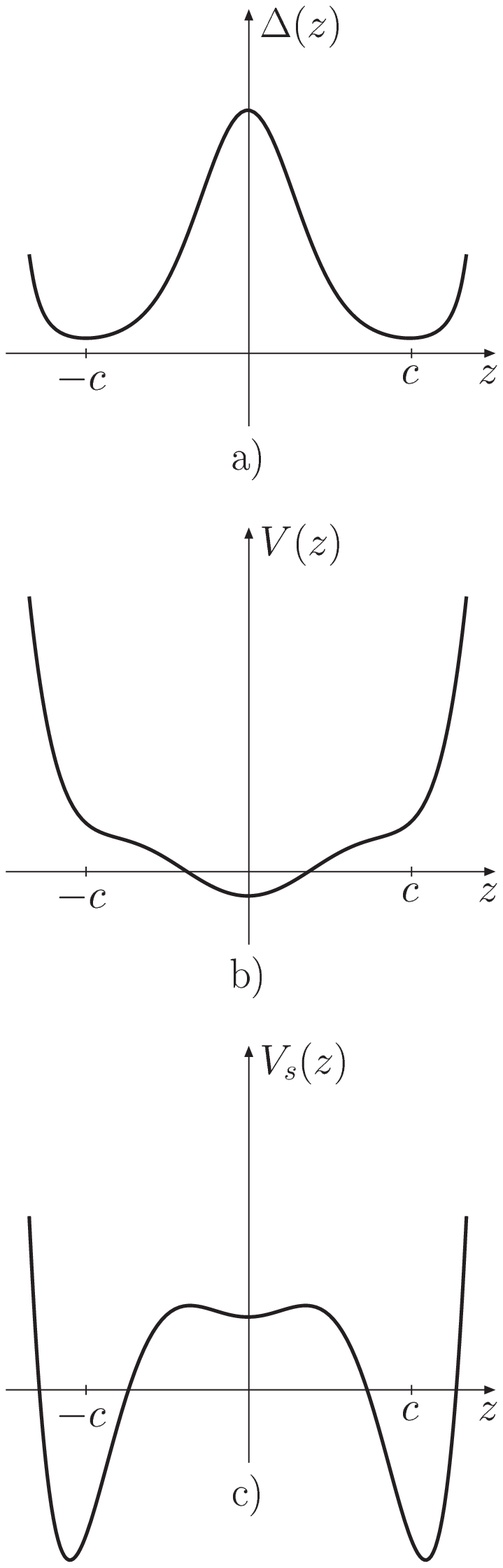,%
bbllx=205pt,bblly=580pt,%
bburx=410pt,bbury=780pt,
width=130pt,height=140pt,%
clip=}}
\put(170,10){
\epsfig{file=fig1.ps,%
bbllx=205pt,bblly=368pt,%
bburx=410pt,bbury=564pt,
width=130pt,height=140pt,%
clip=}}
\put(335,-6){
\epsfig{file=fig1.ps,%
bbllx=205pt,bblly=136pt,%
bburx=410pt,bbury=350pt,
width=130pt,height=160pt,%
clip=}}
\end{picture}
\caption{The weight function $\Delta(z)$, corresponding
quantum-mechanical potential $V(z)$ and its SUSY-partner potential 
$V_s(z)$.}
\end{figure}

We now note that the potential (\ref{pot}) has the same form as in the
supersymmetric quantum mechanics (for a review see \cite{Cooper}).
Thus, one can immediately construct a SUSY-partner potential, $V_s(z)
= W^2 + W'$, which has the same discrete levels plus an extra one, the
exact zero mode \cite{Cooper}. The lowest level of the original
potential (\ref{pot}) is the first excited state for the SUSY-partner
potential $V_s$.  The SUSY-partner potential corresponding to the
weight function of Fig.~1a is plotted in Fig.~1c. It is of the
double-well type, the wells corresponding to the minima of $\Delta(z)$.
The levels which lie below the barrier separating the two minima come
in pairs with exponentially small splitting. Since we know that the
lowest eigenvalue is exactly zero, the next one (which is, in turn,
the lowest for the original potential (\ref{pot})) is exponentially
small. This smallness is due to the exponential suppression of the
tunneling between the two wells.

As a first specific example consider the weight function of the
form
\be
\Delta(z) = \exp\left( - M|z| + {1\over 2} m^2 z^2\right), ~~ M > 0,
~~m^2 > 0, 
\label{ex1}
\ee 
which corresponds to the potential and SUSY-partner potential 
\begin{eqnarray}
\nonumber
V(z) &=& {1\over 4} \left( M - m^2|z| \right)^2 
+ {1\over 2} m^2 -M \delta(z),\\
\nonumber
V_s(z) &=& {1\over 4} \left( M - m^2|z| \right)^2 
- {1\over 2} m^2 + M \delta(z).
\end{eqnarray}
The lowest level $m_0^2$ is zero to all orders of perturbation theory
in $m^2/M^2$. The simplest way to see this is to note that this
perturbation theory would correspond to expanding the weight function
to some finite order in $m^2$, 
\[
\Delta(z) = \exp(- M|z|)\left(1+ {1\over 2} m^2 z^2 +\ldots\right).
\] 
Since the resulting weight function exponentially decreases at large
$z$, eq.(\ref{main}) has an exact normalizable zero mode
$\psi_0= {\rm const}$ no matter at which order the expansion is
truncated. Thus, perturbative corrections are equal to zero. The
semiclassical calculation of the splitting of the two lowest levels in
the SUSY-partner potential $V_s(z)$ gives, for $M^2\gg m^2$,
\[
m_0^2 \sim M m \exp\left( - {M^2\over 2m^2}\right), 
\]
while higher levels are discrete and have masses of order $m$ and
higher. 

Another example with a smooth weight function is 
\be
\Delta(z) = \exp\left(  - {1\over 2} M^2 z^2 +  
{1\over 4}m^4 z^4\right), ~~ M^2 > 0, ~~m^4 > 0.
\label{ex2}
\ee 
The corresponding potential and SUSY-partner potential are 
\begin{eqnarray}
\nonumber
V(z) &=& {1\over 4}z^2 \left( M^2 - m^4 z^2\right)^2 - {1\over 2} M^2 + 
{3\over 2}  m^4 z^2 ,\\
\nonumber
V_s(z) &=& {1\over 4} z^2\left( M^2 - m^4 z^2\right)^2 + {1\over 2}
M^2 - {3\over 2} m^4 z^2 .
\end{eqnarray}
In this case the lowest eigenvalue is, for $M^2\gg m^2$,
\be
m_0^2 \sim {M^4\over m^2} \exp\left( - {M^4\over 4m^4}\right). 
\label{mass2}
\ee
The higher levels are also discrete and have masses starting from $M$. 

In both examples, the lowest eigenvalue is determined, up to a
numerical coefficient, by the ratio $\Delta(c)/\Delta(0)$, where 
$\Delta(c)$ is the minimum of the weight function (see Fig.1a). This
is not a coincidence. Indeed, if $W'\ll W^2$, as in the above
examples, the leading contribution to the tunneling exponent 
\[
m_0^2 \propto \exp\left( -2\int_0^{z_*} \sqrt{V_s(z)}dz \right) \sim
\exp\left( -2\int_0^{z_*}W(z)dz \right)~,
\]
where $z_*$ is a root of the equation $V_s(z)=0$, is
\[
m_0^2 \propto {\Delta(c)\over \Delta(0)}.
\]
Thus, the exponential smallness of the lowest energy level in the
above examples is related to the exponential form of the weight
function. It may lead to a solution of the gauge hierarchy problem,
provided the weight function of exponential form arises
naturally.

Consider now in more detail the effective theory for light modes in the 
case of a non-abelian gauge field in 5 dimensions. This is a 
4-dimensional theory of massive vector field with the standard cubic
and quartic interactions 
\[
g_3 f^{abc}\partial_{\mu} A^a_{\nu} A^{b\mu}A^{c\nu} + 
{g_4 \over 4} f^{abc}f^{adf}A^b_{\mu}A^c_{\nu}A^{d\mu}A^{f\nu},
\]
where the couplings are expressed in terms of the 5-dimensional gauge
coupling $G$ by the following equations,
\begin{eqnarray}
\label{g3}
g_3 &=& G \int dz \Delta(z) \psi_0^3(z),\\
\label{g4}
g_4 &=& G^2 \int dz \Delta(z) \psi_0^4(z).
\end{eqnarray}
Two cases should be distinguished. In the first one the integral $\int
dz \Delta(z)$ converges, and the theory has a normalizable zero
mode, which does not depend on $z$. In this case an effective
4-dimensional theory is an unbroken gauge theory with $g_4 = g_3^2$,
as is dictated by the 4-dimensional gauge invariance. 

In the opposite case the gauge symmetry in 4d is realized in a
different way. To understand it better we put the system in a finite
box with $-z_0 < z < z_0$, and consider a limit $z_0\rightarrow
\infty$ at the end. Then, the constant wave function
$\bar{\psi}_0=const$ (we put a bar to distinguish it from the lowest
massive mode) is normalizable and is a part of the physical
spectrum.  As in the discussion above, it is orthogonal to other
modes.  The corresponding 4d gauge field $\bar{A}^0_\mu$ has a zero
mass and the effective Lagrangian has 4d gauge invariance, with the
massive fields being the gauge singlets. Now, if $\Delta \rightarrow
\infty$ at $z\to \infty$, one can easily see that this massless 
vector field decouples from the other fields in the limit $z_0 \rightarrow
\infty$ and, therefore, is not observable. Indeed, the gauge coupling
of this field is simply
\be
\bar{g} = G \frac{\int dz \Delta\bar{\psi}_0^3}{\int dz
\Delta\bar{\psi}_0^2}=\frac{G}{\sqrt{\int dz \Delta}} \rightarrow 0
\ee
when $z_0 \rightarrow \infty$. 

As for the self-interaction of the lowest massive mode, it survives
in the limit $z_0 \rightarrow \infty$, but  the relation between
quartic and cubic couplings may acquire corrections. By the same
argument as in the case of $m_0^2$, the ratio $g_3^2/g_4$ cannot have
perturbative  corrections. Thus, in the case of the weight function
(\ref{ex1}) one  expects
\[
g_4 = g_3^2\left(1 + O\left({\Delta(c)\over \Delta(0)}\right)\right).
\]

A few comments are now in order:

(i) If the initial theory is formulated in five or higher dimensions,
it is not renormalizable, so it is not surprising that the low-energy
theory is not renormalizable either. In this case one may hope that
if the higher-dimensional theory is in turn an effective field theory
coming from string theory, the divergences will be cut at higher
energy scale. Moreover, the effective theory is nothing but a 
non-linear gauged sigma-model
\cite{Bardeen:1978cz,Appelquist:1980vg}. This model is known to be
strongly interacting at the energy scales of the order $(64 \pi
m_0^2)/3g_4$ \cite{Lee:1977eg}. 

(ii) An interesting situation arises if we start with a gauge theory
formulated in the space-time of 4 or lower dimensions.  In this case
the initial theory is renormalizable, but the low-energy theory in the
space of lower dimensions is not. Divergences disappear after summing
up all heavy degrees of freedom!

(iii) If the electroweak symmetry breaking is to be attributed to this
mechanism, the $\gamma-Z$ mixing must be explained. In the simple
examples we considered above all vector bosons of a simple gauge group
acquire the same masses, and, since the weight function was a gauge
singlet, the required mixing cannot appear at all.  So, to get a
phenomenologically acceptable situation, the effective weight function
should carry gauge indices as well.  Potentially, that could happen in
Kaluza-Klein type compactifications \cite{Witten} generalized to the
warped metric case, or to non-trivial gauge backgrounds
\cite{Ran,Randjbar}.

(iv) The effective low energy action discussed above does not include
any radiative corrections. It remains to be seen whether they affect
the 4-dimensional character of the effective theory.

In conclusion, we proposed that higher dimensional gauge theories may
lead to low energy theories for massive vector bosons without
fundamental scalar fields. It would be interesting to see whether the
weight factors required can arise in a natural way from warped
compactifications on topological defects or in warped string
compactifications \cite{Chan:2000ms}\footnote{After our paper appeared
in arXiv, a paper, \cite{Dvali:2001qr}, appeared in which spontaneous
symmetry breaking in higher dimensional theories was adressed in a
different manner.}.

We wish to thank S. Dubovsky, T. Gherghetta, M.~Laine and V.A.~Rubakov
for helpful discussions.  This work was supported by the FNRS,
contracts no.  21-55560.98, 7SUPJ62239 and 21-58947.99.


\newpage

\begin{thebibliography}{99}
\bibitem{Cornwall}
J.~M.~Cornwall, D.~N.~Levin and G.~Tiktopoulos,
Phys.\ Rev.\ D {\bf 10} (1974) 1145.

\bibitem{Dvali:1997xe}
G.~Dvali and M.~Shifman,
Phys.\ Lett.\ B {\bf 396} (1997) 64.

\bibitem{baj}
B.~Bajc and G.~Gabadadze,
Phys.\ Lett.\  {\bf B474} (2000) 282.

\bibitem{oda}
I.~Oda,
Phys.\ Lett.\ B {\bf 496} (2000) 113.

\bibitem{Dubovsky}
S.~L.~Dubovsky, V.~A.~Rubakov and P.~G.~Tinyakov,
JHEP{\bf 0008} (2000) 041.

\bibitem{Dvali:2001rx}
G.~Dvali, G.~Gabadadze and M.~Shifman,
Phys.\ Lett.\ B {\bf 497} (2001) 271.

\bibitem{Kehagias}
A.~Kehagias and K.~Tamvakis,
hep-th/0010112.

\bibitem{Dimopoulos:2000ej}
P.~Dimopoulos, K.~Farakos, A.~Kehagias and G.~Koutsoumbas,
hep-th/0007079.

\bibitem{Hosotani}
Y.~Hosotani,
Phys.\ Lett.\ B {\bf 126} (1983) 309.\\
P.~Forgacs and N.~S.~Manton,
Commun.\ Math.\ Phys.\ {\bf 72} (1980) 15.

\bibitem{RS2}
V.~A.~Rubakov and M.~E.~Shaposhnikov,
Phys.\ Lett.\  {\bf B125} (1983) 139.

\bibitem{seif}
S.~Randjbar-Daemi and C.~Wetterich,
Phys.\ Lett.\  {\bf B166} (1986) 65.

\bibitem{rusu}
L.~Randall and R.~Sundrum,
Phys.\ Rev.\ Lett.\  {\bf 83} (1999) 4690.

\bibitem{gs}
T.~Gherghetta and M.~Shaposhnikov,
Phys.\ Rev.\ Lett.\  {\bf 85} (2000) 240.

\bibitem{Cooper}
F.~Cooper, A.~Khare and U.~Sukhatme,
Phys.\ Rept.\ {\bf 251} (1995) 267.

\bibitem{Bardeen:1978cz}
W.~A.~Bardeen and K.~Shizuya,
Phys.\ Rev.\ D {\bf 18} (1978) 1969.

\bibitem{Appelquist:1980vg}
T.~Appelquist and C.~Bernard,
Phys.\ Rev.\ D {\bf 22} (1980) 200.

\bibitem{Lee:1977eg}
B.~W.~Lee, C.~Quigg and H.~B.~Thacker,
Phys.\ Rev.\ D {\bf 16} (1977) 1519.

\bibitem{Witten}
E.~Witten,
Nucl.\ Phys.\ B {\bf 186} (1981) 412.

\bibitem{Ran}
S.~Randjbar-Daemi, A.~Salam and J.~Strathdee,
Nucl.\ Phys.\ B {\bf 214} (1983) 491.

\bibitem{Randjbar}
S.~Randjbar-Daemi and M.~Shaposhnikov,
Phys.\ Lett.\ B {\bf 491} (2000) 329

\bibitem{Chan:2000ms}
C.~S.~Chan, P.~L.~Paul and H.~Verlinde,
Nucl.\ Phys.\ B {\bf 581} (2000) 156.

\bibitem{Dvali:2001qr}
G.~Dvali, S.~Randjbar-Daemi and R.~Tabbash,
hep-ph/0102307.

\end{thebibliography}
\end{document}